\documentclass[ALICE,manyauthors]{cernphprep}
\usepackage[comma,square,numbers,sort&compress]{natbib}

\usepackage{lineno}
\modulolinenumbers[5]

\usepackage{xspace}
\usepackage{hyperref}

\usepackage[dvipsnames]{xcolor}

\usepackage{enumerate}
\usepackage{comment}
\usepackage{bbold} 
\usepackage{booktabs} 
\usepackage{multirow}

\usepackage[T1]{fontenc}
\usepackage{orcidlink}

\begin{document}
%

\newcommand{\pp}           {pp\xspace}
\newcommand{\ppbar}        {\mbox{$\mathrm {p\overline{p}}$}\xspace}
\newcommand{\XeXe}         {\mbox{Xe--Xe}\xspace}
\newcommand{\PbPb}         {\mbox{Pb--Pb}\xspace}
\newcommand{\pA}           {\mbox{pA}\xspace}
\newcommand{\pPb}          {\mbox{p--Pb}\xspace}
\newcommand{\AuAu}         {\mbox{Au--Au}\xspace}
\newcommand{\dAu}          {\mbox{d--Au}\xspace}

\newcommand{\s}            {\ensuremath{\sqrt{s}}\xspace}
\newcommand{\snn}          {\ensuremath{\sqrt{s_{\mathrm{NN}}}}\xspace}
\newcommand{\pt}           {\ensuremath{p_{\rm T}}\xspace}
\newcommand{\meanpt}       {$\langle p_{\mathrm{T}}\rangle$\xspace}
\newcommand{\ycms}         {\ensuremath{y_{\rm CMS}}\xspace}
\newcommand{\ylab}         {\ensuremath{y_{\rm lab}}\xspace}
\newcommand{\etarange}[1]  {\mbox{$\left | \eta \right |~<~#1$}}
\newcommand{\yrange}[1]    {\mbox{$\left | y \right |~<~#1$}}
\newcommand{\dndy}         {\ensuremath{\mathrm{d}N_\mathrm{ch}/\mathrm{d}y}\xspace}
\newcommand{\dndeta}       {\ensuremath{\mathrm{d}N_\mathrm{ch}/\mathrm{d}\eta}\xspace}
\newcommand{\avdndeta}     {\ensuremath{\langle\dndeta\rangle}\xspace}
\newcommand{\dNdy}         {\ensuremath{\mathrm{d}N_\mathrm{ch}/\mathrm{d}y}\xspace}
\newcommand{\Npart}        {\ensuremath{N_\mathrm{part}}\xspace}
\newcommand{\Ncoll}        {\ensuremath{N_\mathrm{coll}}\xspace}
\newcommand{\dEdx}         {\ensuremath{\textrm{d}E/\textrm{d}x}\xspace}
\newcommand{\RpPb}         {\ensuremath{R_{\rm pPb}}\xspace}

\newcommand{\nineH}        {$\sqrt{s}~=~0.9$~Te\kern-.1emV\xspace}
\newcommand{\seven}        {$\sqrt{s}~=~7$~Te\kern-.1emV\xspace}
\newcommand{\twoH}         {$\sqrt{s}~=~0.2$~Te\kern-.1emV\xspace}
\newcommand{\twosevensix}  {$\sqrt{s}~=~2.76$~Te\kern-.1emV\xspace}
\newcommand{\five}         {$\sqrt{s}~=~5.02$~Te\kern-.1emV\xspace}
\newcommand{\twosevensixnn}{$\sqrt{s_{\mathrm{NN}}}~=~2.76$~Te\kern-.1emV\xspace}
\newcommand{\fivenn}       {$\sqrt{s_{\mathrm{NN}}}~=~5.02$~Te\kern-.1emV\xspace}
\newcommand{\twohundrednn}{$\sqrt{s_{\mathrm{NN}}}~=~200$~Ge\kern-.1emV\xspace}
\newcommand{\hundrednintythreenn}{$\sqrt{s_{\mathrm{NN}}}~=~193$~Ge\kern-.1emV\xspace}
\newcommand{\fivefourfournn}       {$\sqrt{s_{\mathrm{NN}}}~=~5.44$~Te\kern-.1emV\xspace}

\newcommand{\LT}           {L{\'e}vy-Tsallis\xspace}
\newcommand{\GeVc}         {Ge\kern-.1emV/$c$\xspace}
\newcommand{\MeVc}         {Me\kern-.1emV/$c$\xspace}
\newcommand{\TeV}          {Te\kern-.1emV\xspace}
\newcommand{\GeV}          {Ge\kern-.1emV\xspace}
\newcommand{\MeV}          {Me\kern-.1emV\xspace}
\newcommand{\GeVmass}      {Ge\kern-.2emV/$c^2$\xspace}
\newcommand{\MeVmass}      {Me\kern-.2emV/$c^2$\xspace}
\newcommand{\lumi}         {\ensuremath{\mathcal{L}}\xspace}

\newcommand{\ITS}          {\rm{ITS}\xspace}
\newcommand{\TOF}          {\rm{TOF}\xspace}
\newcommand{\ZDC}          {\rm{ZDC}\xspace}
\newcommand{\ZDCs}         {\rm{ZDCs}\xspace}
\newcommand{\ZNA}          {\rm{ZNA}\xspace}
\newcommand{\ZNC}          {\rm{ZNC}\xspace}
\newcommand{\SPD}          {\rm{SPD}\xspace}
\newcommand{\SDD}          {\rm{SDD}\xspace}
\newcommand{\SSD}          {\rm{SSD}\xspace}
\newcommand{\TPC}          {\rm{TPC}\xspace}
\newcommand{\TRD}          {\rm{TRD}\xspace}
\newcommand{\VZERO}        {\rm{V0}\xspace}
\newcommand{\VZEROA}       {\rm{V0A}\xspace}
\newcommand{\VZEROC}       {\rm{V0C}\xspace}
\newcommand{\Vdecay} 	   {\ensuremath{V^{0}}\xspace}

\newcommand{\ee}           {\ensuremath{e^{+}e^{-}}} 
\newcommand{\pip}          {\ensuremath{\pi^{+}}\xspace}
\newcommand{\pim}          {\ensuremath{\pi^{-}}\xspace}
\newcommand{\kap}          {\ensuremath{\rm{K}^{+}}\xspace}
\newcommand{\kam}          {\ensuremath{\rm{K}^{-}}\xspace}
\newcommand{\pbar}         {\ensuremath{\rm\overline{p}}\xspace}
\newcommand{\kzero}        {\ensuremath{{\rm K}^{0}_{\rm{S}}}\xspace}
\newcommand{\lmb}          {\ensuremath{\Lambda}\xspace}
\newcommand{\almb}         {\ensuremath{\overline{\Lambda}}\xspace}
\newcommand{\Om}           {\ensuremath{\Omega^-}\xspace}
\newcommand{\Mo}           {\ensuremath{\overline{\Omega}^+}\xspace}
\newcommand{\X}            {\ensuremath{\Xi^-}\xspace}
\newcommand{\Ix}           {\ensuremath{\overline{\Xi}^+}\xspace}
\newcommand{\Xis}          {\ensuremath{\Xi^{\pm}}\xspace}
\newcommand{\Oms}          {\ensuremath{\Omega^{\pm}}\xspace}
\newcommand{\degree}       {\ensuremath{^{\rm o}}\xspace}

\newcommand{\rz}         {\ensuremath{\rho^{0}}\xspace}
\newcommand{\Psip}         {\ensuremath{\psi^\prime}\xspace}
\newcommand{\Noon}         {\textbf{n$\mathbf{_O^O}$n}}
\newcommand{\WgPb}{\ensuremath{W_{\gamma\, \mathrm{Pb, n}}}\xspace}
\newcommand{\gPb}{\gamma\,\mathrm{Pb}}
\newcommand{\ptdimuon}     {\ensuremath{p^{\mu\mu}_{\rm T}}\xspace}
\newcommand{\mdimuon}     {\ensuremath{m_{\mu\mu}}\xspace}
\newcommand{\pttwo}        {\ensuremath{p^{2}_{\rm T}}\xspace}
\newcommand{\mant}         {\ensuremath{|t|}\xspace}
\newcommand{\ggll}         {\ensuremath{\gamma\gamma\to l^+ l^-}\xspace}

\newcommand{\BR}           {\ensuremath{\rm BR}\xspace}
\newcommand{\axe}          {\ensuremath{\rm{Acc\times\epsilon}}\xspace}
\newcommand{\fC}           {\ensuremath{f_{\rm C}}\xspace}
\newcommand{\fD}           {\ensuremath{f_{\rm D}}\xspace}

\newcommand{\AD}          {\rm{AD}\xspace}
\newcommand{\IP}          {\rm{IP}\xspace}
\newcommand{\rhoz}          {\ensuremath{\rho^{0}}\xspace}
\newcommand{\GeVcSqr}         {Ge\kern-.1emV/$c^2$\xspace}
\newcommand{\MeVcSqr}         {Me\kern-.1emV/$c^2$\xspace}
\newcommand{\mpipi}        {\ensuremath{m_{\pi\pi}}\xspace}
\newcommand{\mpipiEq}[1]        {\ensuremath{m_{\pi\pi}=#1}\xspace}
\newcommand{\GeVcq}         {$($Ge\kern-.1emV/$c)^2$\xspace}
\newcommand{\phiz}          {\ensuremath{\phi}\xspace}
\newcommand{\UU}         {\mbox{U--U}\xspace}
\newcommand{\starTwo}       {$\sqrt{s_{\mathrm{NN}}}~=~200$~Ge\kern-.1emV\xspace}
\newcommand{\etal}   {\textit{et al.}}
\newcommand{\add}{\colorbox{Orange}{\textcolor{white}{To add}}}
\newcommand{\tw}{\colorbox{Red}{\textcolor{white}{To write}}}
\newcommand{\wrote}{\colorbox{Blue}{\textcolor{white}{Written}}}
\newcommand{\ck}{\colorbox{Purple}{\textcolor{white}{To check}}}

\begin{titlepage}
\PHyear{2024}       
\PHnumber{126}      
\PHdate{13 May}     

\title{
Measurement of the impact-parameter dependent azimuthal anisotropy in coherent $\mathbf{\rho^0}$ photoproduction in Pb--Pb collisions at $\mathbf{\sqrt{s_{\rm{NN}}}~=~5.02}~\textbf{TeV}$}
\ShortTitle{Azimuthal anisotropy in coherent $\mathbf{\rho^0}$ photoproduction}   

\Collaboration{ALICE Collaboration\thanks{See Appendix~\ref{app:collab} for the list of collaboration members}}
\ShortAuthor{ALICE Collaboration} 

\begin{abstract}
This Letter presents the first measurement of the impact-parameter dependent angular anisotropy in the decay of coherently photoproduced $\rho^0$ mesons. The \rhoz mesons are reconstructed through their decay into pion pairs. The measured anisotropy corresponds to the amplitude of the $\cos(2\phi)$ modulation, where $\phi$ is the angle between the two vectors formed by the sum and  the difference of the transverse momenta of the pions, respectively. The measurement was performed by the ALICE Collaboration at the LHC using  data from ultraperipheral \PbPb collisions at a center-of-mass energy of \fivenn per nucleon pair.
Different  impact-parameter regions are selected by classifying the events in nuclear-breakup classes.
The amplitude of the $\cos(2\phi)$ modulation is found to increase by about one order of magnitude  from large to small impact parameters.
Theoretical calculations describe the measured $\cos(2\phi)$ anisotropy and its impact-parameter dependence as the result of a quantum interference effect at the femtometer scale, arising from the ambiguity regarding which of the nuclei is the photon source in the interaction.
\end{abstract}

\end{titlepage}

\setcounter{page}{2} 


\section{Introduction
\label{sec:intro}}

The heavy ions circulating at the Relativistic Heavy-Ion Collider (RHIC) and Large Hadron Collider (LHC) accelerators are accompanied by a strong, Lorentz contracted, electromagnetic field that can be described as a flux of quasi-real photons. This flux makes it possible to study photoproduction interactions at these facilities~\cite{Baur:2001jj,Baltz:2007kq,Contreras:2015dqa,Klein:2019qfb}. Most experimental work uses ultraperipheral collisions (UPCs), where the colliding ions cross paths at impact parameters larger than the sum of their radii. Given the short range of the strong force, UPCs allow one to separate photon-induced processes from hadronic interactions.  

One of the processes that has received great interest is the  photonuclear production of a vector meson, where the incoming photon fluctuates into a quark--antiquark color dipole that scatters off the nucleus traveling in the opposite direction (denoted as the target), and appears as a real vector meson. This process can be either coherent, if the photon couples to the nucleus as a whole, or incoherent, if it interacts with a single nucleon. The two processes result in a different transverse momentum (\pt) spectrum; the mean \pt of the vector meson is related to the size of the target in the impact-parameter plane and it is around 60 (500) \MeVc in the coherent (incoherent) case. In the coherent scenario, it is not known which of the two colliding nuclei emits the photon and which acts as the target, opening up the possibility  to study, at femtometer scales,  the fundamental quantum mechanical interference between the amplitudes. This idea was first proposed in Ref.~\cite{Klein:1999gv}, where it was noted that interference effects should be stronger: ($i$) around midrapidity, where the magnitude of both amplitudes is similar, and ($ii$)
at small impact parameters $b$, where $b$ acts analogously to the distance between slits, in a two-slit interferometer.  

Coherent vector meson photoproduction accompanied by electromagnetic dissociation (EMD) offers the opportunity to select different impact parameter regions in UPCs~\cite{Baltz:2002pp}.  The electromagnetic field of the heavy ions is so intense that there is a non-negligible  probability that the two nuclei, besides interacting to produce the $\rhoz$, also exchange photons in an independent EMD interaction, where the excited nuclei emit neutrons at beam rapidities. Experimentally,  the emitted neutrons can be detected using two zero-degree calorimeters (ZDCs) each of them covering the direction of one of the incoming colliding nuclei. This allows for classifying UPCs as: (\textit{i}) XnXn, where at least one neutron is detected in each ZDC, (\textit{ii}) Xn0n + 0nXn, where at least one neutron is detected in only one of the ZDCs, and (\textit{iii}) 0n0n, where no neutron is detected in the ZDCs; for brevity,  the class Xn0n + 0nXn will be denoted as Xn0n in the following text. Since the intensity of the electromagnetic field grows with decreasing impact parameter, the XnXn configuration, where at least three photons are exchanged, selects a region of relatively small impact parameters. The 0nXn and Xn0n configurations select a broader impact-parameter range than XnXn, while 0n0n events encompass all possible impact parameters. EMD is modeled in the RELDIS~\cite{Pshenichnov:2001qd,Pshenichnov:2011zz} and \Noon~\cite{Broz:2019kpl} models, while the coherent production of vector mesons accompanied by electromagnetic dissociation is studied with \Noon\  and STARlight~\cite{Klein:2016yzr}. According to \Noon, the median impact parameter of coherent \rhoz photoproduction at the LHC energy changes from about 49~fm in 0n0n to about 19~fm in XnXn.

Ref.~\cite{Klein:1999gv} proposes the suppression of coherent $\rhoz$ production at small transverse momentum in UPCs as an observable to study interference effects.  This effect was measured  by the STAR Collaboration in coherent $\rhoz$ photoproduction at a center-of-mass energy per nucleon pair of \twohundrednn~\cite{STAR:2008llz}. The measurement was carried out using two samples, one corresponding to XnXn and the other without any requirement on the detection of neutrons at beam rapidities. As expected, it was observed that the destructive interference was more pronounced in the XnXn sample.

Recently, it has been pointed out that interference can also give rise to azimuthal anisotropy, since the incoming photons are linearly polarized. It was suggested to look for this effect in the process $\gamma+\gamma\to l^++l^-$ where $\gamma$ and $l^{\pm}$ denote  photons and leptons, respectively~\cite{Li:2019yzy}. The dependence of this phenomenon on the impact parameter was studied in Ref.~\cite{Li:2019sin}. Shortly thereafter, this effect was measured for the XnXn event class by the STAR Collaboration in \AuAu UPCs at \twohundrednn~\cite{STAR:2019wlg}.

These studies were later extended to the photoproduction of a $\rhoz$ vector meson, where the $\rhoz$ inherits the linear polarization of the photon, giving rise to a  $\cos(2\phi)$ asymmetry~\cite{Xing:2020hwh,Zha:2020cst}. Here, $\phi$ is the angle between the two vectors formed by the sum and by the difference of the transverse momenta of the pions produced in the decay $\rhoz\to\pi^+\pi^-$. More recently, it was proposed to look for $\cos(\phi)$, $\cos(3\phi)$~\cite{Hagiwara:2020juc}, and  $\cos(4\phi)$~\cite{Hagiwara:2021xkf} modulations. The first two patterns could be produced by the interference of the production of $\rhoz$ with QED processes, and the last  by the interference of resonant and open production of pion pairs. It was also proposed  to search for asymmetries in the photoproduction of a J$/\psi$ vector meson~\cite{Brandenburg:2022jgr}.
The predicted $\cos(2\phi)$ modulation was measured by the STAR Collaboration, for $\rhoz$ coherent production in XnXn events, in \AuAu and \UU UPCs at \twohundrednn and \hundrednintythreenn, respectively~\cite{STAR:2022wfe}. This asymmetry was also recently studied by the CMS Collaboration using exclusive diffractive production of jets at the LHC~\cite{CMS:2022lbi}.

The ALICE Collaboration has measured the cross section for EMD in Pb--Pb collisions at center-of-mass energies of \twosevensixnn ~\cite{ALICE:2012aa} and \fivenn~\cite{ALICE:2022iqi}, where a good agreement with the predictions from RELDIS and \Noon\ was found. The ALICE Collaboration has also measured coherent $\rhoz$ photoproduction in \PbPb UPCs at \twosevensixnn~\cite{ALICE:2015nbw} and  \fivenn~\cite{ALICE:2020ugp}, as well as in \XeXe UPCs at \fivefourfournn~\cite{ALICE:2021jnv}. The results were compared to predictions from the STARlight and GDL~\cite{Rebyakova:2011vf} models in Ref.~\cite{ALICE:2015nbw}, and  from STARlight and \Noon~in Refs.~\cite{ALICE:2020ugp,ALICE:2021jnv}; in general, the tested models describe well the relative \rhoz yields in the 0n0n, Xn0n, and XnXn classes. These measurements demonstrate that coherent $\rhoz$ photoproduction accompanied by EMD is well understood at the LHC and that neutron emission in EMD can be used to select different event classes which are dominated by different impact parameter ranges.

In this Letter, the impact-parameter dependence of the $\cos(2\phi)$ asymmetry  is studied in \PbPb UPCs at \fivenn using the coherent photoproduction of a $\rhoz$ meson decaying into a pion pair. The measurements are performed at midrapidity in the range $|y|<0.8$ and in three different EMD classes: 0n0n, Xn0n, and XnXn.

\section{Experimental set-up
\label{sec:setup}}

A full description of the ALICE apparatus and its performance is given in Refs.~\cite{Aamodt:2008zz,Abelev:2014ffa}. A brief description of the sub-detectors involved in this analysis is given hereafter. The \rhoz meson is detected through its decay into a pion pair at midrapidity, using the Inner Tracking System (\ITS)~\cite{Aamodt:2010aa} and the Time Projection Chamber (\TPC)~\cite{Alme:2010ke} to reconstruct the pion tracks. The \VZERO~\cite{Abbas:2013taa} and ALICE Diffractive (\AD)~\cite{LHCForwardPhysicsWorkingGroup:2016ote} detectors, located at forward rapidities, provide a veto,  suppressing hadronic interactions. As mentioned in Sec.~\ref{sec:intro}, three different impact-parameter ranges are selected by means of neutrons emitted at forward rapidities, measured by the ZDCs~\cite{ALICE:2012aa}.

The \ITS comprises six cylindrical layers coaxial with the beam line. Three different technologies are used, starting from the inner layer: pixel, drift, and strip sensors. Each technology is used in two consecutive layers. All six layers are used for tracking, while the two innermost layers, the Silicon Pixel Detector (\SPD),  are also used for triggering.  The TPC is a large cylindrical gaseous detector that surrounds the \ITS. It has a central cathode at high voltage and two readout planes at the end caps, composed of multiwire proportional chambers. It is the main tracking detector and provides particle identification (PID) by measuring the specific ionization energy loss. The \ITS and the \TPC cover a pseudorapidity interval $|\eta|<0.9$ and the full azimuth; they are located inside a solenoid magnet that provides a magnetic field of $B = 0.5$ T.

The \VZERO is composed of two scintillator arrays, \VZEROA and \VZEROC, installed on both sides of the nominal interaction point (\IP). They cover the pseudorapidity ranges  \mbox{$2.8 < \eta < 5.1$} and \mbox{$-3.7 < \eta < -1.7$}, respectively. The \AD consists of two scintillator stations, ADA and ADC, located along the beam line at $+16$~m and $-19$~m from the \IP and covering the pseudorapidity ranges \mbox{$4.8 < \eta < 6.3$} and \mbox{$-7.0 < \eta < -4.9$}, respectively.

There are two \ZDC detectors for neutrons, \ZNA and \ZNC, located at $\pm 112.5$~m from the \IP along the beam line. They detect neutrons with $|\eta|>8.8$, with an energy resolution good enough to be sensitive to the emission of a single neutron. Neutron signals are discriminated with a threshold corresponding to an energy deposition of $\sim$~1~TeV, which is about three standard deviations below the expected signal from a 2.51~TeV neutron. The ZNs also determine the arrival time of the particles, allowing for the rejection of beam--gas interactions involving charge circulating outside the nominal LHC bunch positions.

The analyzed data were recorded by ALICE in 2015, when the LHC provided \PbPb collisions at \fivenn, using a dedicated UPC trigger. This trigger exploits five different signals: four of them veto any activity on either side of the \AD or \VZERO detector within the time window for nominal beam--beam interactions, to suppress hadronic collisions. The fifth signal is a topological trigger that selects events that have at least two track segments, defined as in Ref.~\cite{ALICE:2020ugp}, in the \SPD, with an opening angle in azimuth greater than 153 degrees. This topology was chosen since the coherently produced \rhoz has a very small transverse momentum and hence the tracks of the pions are almost back-to-back in azimuth. The integrated luminosity of the sample, determined using the \VZERO detectors as explained in Ref.~\cite{ALICE:2020ugp}, is about 485~mb$^{-1}$.

\section{Analysis procedure
\label{sec:ana}}

\subsection{Track and event selection \label{sec:evSelection}}
Tracks were required to have a distance of closest approach to the event primary vertex smaller than $0.0182~+~0.0350/(p_\text{T}^{\text{trk}})^{1.01}$~cm in the transverse plane and smaller than 2~cm in the longitudinal direction, where $p_\text{T}^{\text{trk}}$ is the transverse momentum, in \GeVc, associated to the track. Tracks were also required to have more than 50 associated hits in the \TPC, to be  reconstructed in both \ITS and \TPC, and to match the track segments in the \SPD that fired the trigger.

The events with good tracks were required to fulfill additional selections: (\textit{i})~have exactly two tracks of opposite sign, (\textit{ii})~have no offline reconstructed signal in neither the \VZERO nor \AD detectors, and (\textit{iii})~fulfill the pion selection $\text{n}_{\sigma 1}^2 + \text{n}_{\sigma 2}^2 < 5^2$, where $\text{n}_{\sigma 1}$ ($\text{n}_{\sigma 2}$) is the difference, in units of the TPC ionization energy loss resolution, between the measured energy loss for track 1 (track 2) and the expected value for a pion with the same momentum. 

Kinematic selections were also applied: 
(\textit{i})~the pion pair rapidity lies in the range $|y|<0.8$ to avoid acceptance edge effects, (\textit{ii})~the invariant mass of the pion pair is inside the range $0.6~\text{\GeVcSqr}<\text{\mpipi}~<0.95 \text{ \GeVcSqr}$, and (\textit{iii})~the transverse momentum of the \rhoz candidate is less than 0.1~\GeVc to select coherent processes with high purity. With such a selection, the contamination from incoherent events is found to be lower than 4\%~\cite{ALICE:2020ugp}.
More details about event and track selections can be found in Ref.~\cite{ALICE:2020ugp}.

The data were divided in three independent classes, based on the detection of neutrons at forward rapidity. Events with neutron emission were selected by requiring a signal in ZNA and/or ZNC. The signal time was required to lie within 2~ns from the nominal collision time.
As explained in Sec.~\ref{sec:intro}, these classes (0n0n, Xn0n, XnXn) can be used to select different impact-parameter ranges for ultraperipheral collisions. For each neutron class, the data were arranged in seven $\phi$ intervals of equal size, with $\phi$ defined as in Sec.~\ref{sec:phiDef}. The efficiency correction (Sec.~\ref{Sec:corrections}) and signal extraction (Sec.~\ref{Sec:signalExtraction}) were then performed separately for each neutron class and $\phi$ interval.

\subsection{Asymmetry angle definition \label{sec:phiDef} }
The anisotropy described in Sec.~\ref{sec:intro} is predicted to be strongly visible as a function of a variable called $\phi$, defined using the momenta of the pions ($\vec{\pi}_{1,2}$) into which the \rhoz decays. The $\phi$ angle can be defined in two different ways, which will be referred to as \textit{average} and \textit{charge}, respectively. In both cases $\phi$ is defined as the angle between the transverse components of $\vec{p}_+$ and $\vec{p}_-$, where $\vec{p}_{\pm} = \vec{\pi}_1 \pm \vec{\pi}_2$. Using the  \textit{average} definition, $\vec{\pi}_{1,2}$ are randomly associated to the positive or to the negative track. Using the \textit{charge} definition, $\vec{\pi}_{1}$ and $\vec{\pi}_{2}$ are the momenta of the positive and of the negative track, respectively. The \textit{average} definition is helpful since, by construction, it does not allow for a $\cos(\phi)$ component. The two definitions are equivalent in terms of the predicted cos(2$\phi$) component. The \textit{average} definition was chosen as the default one, while the \textit{charge} definition was used in the evaluation of the systematic uncertainties. In both cases, the $\phi$ angle was initially computed within the range of $-\pi$ to $\pi$. Then, since cos($\phi$) and cos(2$\phi$) are even functions, the resulting values were remapped between 0 and $\pi$ by flipping the sign of negative values. This procedure improves the stability of signal extraction when fitting the invariant mass spectra in each $\phi$ interval, by doubling the size of the available sample.

\subsection{Corrections \label{Sec:corrections}}
The correction for the acceptance and efficiency of the detector for the reconstruction and selection of the pion tracks, \axe, was estimated as a function of the pion pair invariant mass using the STARlight~\cite{Klein:2016yzr} Monte Carlo (MC) generator and a realistic description of the ALICE apparatus. This MC production describes accurately the raw data on the vector meson kinematics, with the exception of the transverse momentum distribution~\cite{ALICE:2015nbw}. In order to improve the agreement of the MC with data, a re-weighting procedure, described in the following, was applied to the generated \pttwo spectrum. The first step of the procedure is to fit the inclusive pion pair \pttwo distribution of the generated MC.
For sufficiently high transverse momentum, \pttwo can be approximated with the Mandelstam variable $t$, while for very low \pttwo the contribution from the transverse momentum of the photon plays an important role and hence the approximation is no longer valid. Having this in mind,  the MC spectrum was fitted for $\pttwo>(0.01)^2$~\GeVcq using the function
\begin{equation}
    \frac{\text{d}N}{\text{d\pttwo}} = c \mid F(\mid t \mid,~a_{\text{Pb}},~R_{\text{Pb}}) \mid^2,
\end{equation}
where $c$ is a normalization constant and $F(\mid t \mid)$ is the form factor of the lead nucleus, obtained as a numerical approximation of the Fourier transform of a Wood--Saxon function~\cite{Klein:1999qj, Maximon:1966sqn}, with fit parameters $R_{\rm{Pb}}$ and $a_{\rm{Pb}}$. The weights were then computed using
\begin{equation}
    \text{w(\pt)} = \frac{\mid F(\mid t \mid,~a_{\text{Pb}}^0,~R_{\text{X}}) \mid^2}{\mid F(\mid t \mid, ~a_{\text{Pb}}^0,~ R_{\text{Pb}}^0) \mid^2},
    \label{Eq:weight}
\end{equation}
where $ a_{\text{Pb}}^0$ and $R_{\text{Pb}}^0$  are the parameters extracted from the fit to the MC spectrum and $R_{\text{X}}$ is chosen in such a way that, after applying the weights to each event of a given generated \pt, the reconstructed \pttwo spectrum in the MC best reproduces the one in the data. This is achieved by minimizing the bin-by-bin difference between the \pt distributions of data and reconstructed MC as a function of $R_{\text{X}}$, using a $\chi^2$-like variable, in the region $\pttwo>(0.01)^2$ \GeVcq, where the model used for the reweighting is valid. 
It was verified that the best-fit values of $R_{\text{X}}$ computed for different $\phi$ ranges were all compatible among themselves, hence the weights were obtained utilizing the full data set. 

The \axe correction was obtained using the re-weighted STARlight MC simulations by computing  the ratio of reconstructed to generated number of pion pairs in each invariant mass and $\phi$ interval, after applying the weights discussed above at the  generation level. The number of pion pairs found in data, for each invariant mass and $\phi$ interval, was then divided by \axe, to obtain the corrected mass spectra.  This was done for each neutron class and $\phi$ range.
The integrated \axe was found to slightly increase as a function of the invariant mass, ranging from $\sim 10.5\%$ at \mpipiEq0.6 \GeVcSqr to $\sim 14\%$ at \mpipiEq0.95 \GeVcSqr.

\subsection{Signal extraction \label{Sec:signalExtraction}}
The corrected invariant mass spectra, in each neutron class and in each \phiz interval, were fitted with: 
\begin{equation}
    \frac{\text{d}N}{\text{d}m_{\pi\pi}} = P(m_{\pi\pi}) + n_{\mu\mu} \: M(m_{\pi\pi}),
    \label{Eq:fit}
\end{equation}
where $m_{\pi\pi}$ is the pion pair invariant mass, $P(m_{\pi\pi})$ is the function used to describe the pion pair spectrum, $M(m_{\pi\pi})$ is the shape, estimated with a dedicated MC based on the STARlight generator, of the  background originating from muons produced in the $\gamma\gamma \rightarrow \mu^+\mu^-$ process and misidentified as pions, and $n_{\mu\mu}$ is a normalization constant for said background. Two different parameterizations were used for the pion spectrum. The first parameterization uses
a  modified S\"{o}ding model~\cite{PhysRev.159.1496}:
\begin{equation}
    P(m_{\pi\pi}) = \mid A \cdot BW_{\rho} + B \mid^2,
    \label{Eq:soed_full}
\end{equation}
where $BW_{\rho}$ is the relativistic Breit--Wigner shape that describes the \rhoz, $A$ is its amplitude, and $B$ is the amplitude of the continuum pion pair production. The relativistic Breit--Wigner function describing the $\rho^0$ resonance is:
\begin{equation}
    BW_{\rho} = \frac{\sqrt{m_{\pi\pi} \, m_{\rho} \, \Gamma_{\rho}(m_{\pi\pi})}}{m_{\pi\pi}^2-m_{\rho}^2 + i \, m_{\rho} \, \Gamma_{\rho}(m_{\pi\pi})},
\end{equation}
where $m_{\rho}$ is the \rhoz pole mass and
\begin{equation}
    \Gamma_{\rho}(m_{\pi\pi}) = \Gamma(m_{\rho}) \, \frac{m_{\rho}}{m_{\pi\pi}} \, \bigg(\frac{m_{\pi\pi}^2-4m_{\pi}^2}{m_{\rho}^2-4m_{\pi}^2}\bigg)^{3/2},
\end{equation}
where $\Gamma$($m_{\rho}$) is the \rhoz pole width. Since $BW_{\rho}$ is complex, an
interference term between the \rhoz and the continuum arises from the square module in Eq.~(\ref{Eq:soed_full}). The second parameterization uses the model by Ross and Stodolsky~\cite{PhysRev.149.1172}:
\begin{equation}
    P(m_{\pi\pi}) = f  \mid BW_{\rho} \mid^2  \big(\frac{m_{\rho}}{m_{\pi\pi}} \big)^k,
    \label{Eq:ross}
\end{equation}
with fit parameters $f$ and $k$.

The fits were performed by fixing $m_{\rho}$ and $\Gamma(m_{\rho})$ to the central values reported for a \rhoz formed in a photoproduction reaction, namely, $m_{\rho}$~=~769.2~\MeVcSqr and $\Gamma(m_{\rho})$~=~151.5~\MeVcSqr~\cite{ParticleDataGroup:2022pth}. 
As discussed in Ref.~\cite{ALICE:2020ugp}, it was verified that the extracted \rhoz yield does not vary significantly when the fit function is modified to include a contribution from $\omega$ decays.
\newline 

\begin{figure}[!ht]
\centering
\includegraphics[width=0.98\textwidth]{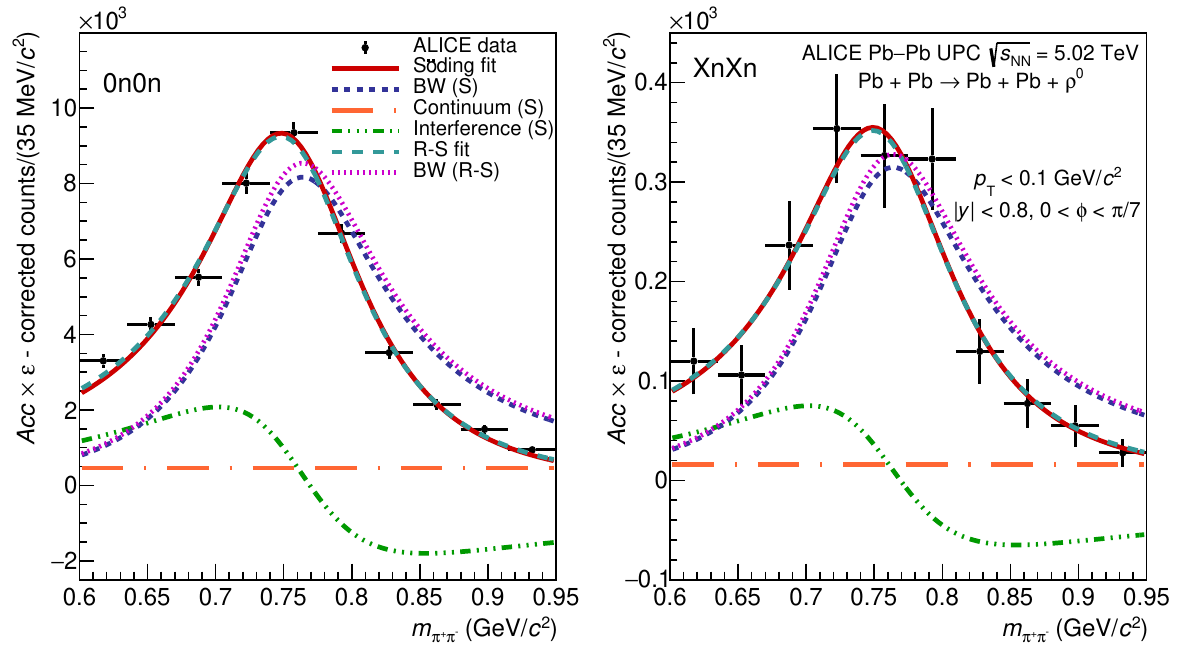}
\caption{Invariant-mass distribution of pion pairs, with superimposed S\"{o}ding (solid line) and Ross-Stodolsky (dotted line) fits, for the range $0<\phi<\pi/7$ in the 0n0n (left) and XnXn (right) neutron classes. The different components of the pion-pair production amplitude in the S\"{o}ding model are shown: the Breit--Wigner shape that describes the \rhoz (finer dotted line), the continuum process (dash-dotted line), and the interference between the \rhoz and the continuum (dash-dot-dot-dot line). The Breit--Wigner extracted from the Ross-Stodolsky model (finest dotted line) is also shown. In this example, the background contribution from misidentified muons is fixed to zero in the fit.}
\label{Fig:massFits}  
\end{figure}

To assess the stability of the signal extraction, it was carried out for each $\phi$ interval using 48 different strategies. These strategies included 12 different combinations of bin size and fit range (with the widest tested range  being 0.6-0.95~\GeVcSqr and the narrowest 0.65-0.9~\GeVcSqr), using the S\"{o}ding or the Ross-Stodolsky model to describe the pion pair mass spectrum, and using $n_{\mu\mu}$ as a free fit parameter or fixing it to zero.   
An example of the mass fits, for a specific  $\phi$ interval and for the 0n0n and XnXn classes, is shown in Fig.~\ref{Fig:massFits}. After the fit, the \rhoz yield is obtained by integrating the signal function, $|A\: BW_{\rho}|^2$ for the S\"oding model and $f|BW_{\rho}|^2$ for the Ross-Stodolsky model, in the mass range $0.6<\text{\mpipi}\enspace(\text{\GeVcSqr}) <0.95$. 

\subsection{Asymmetry extraction \label{Sec:asymmExtraction}}
\begin{figure}[!ht]
\centering
\includegraphics[width=0.98\textwidth]{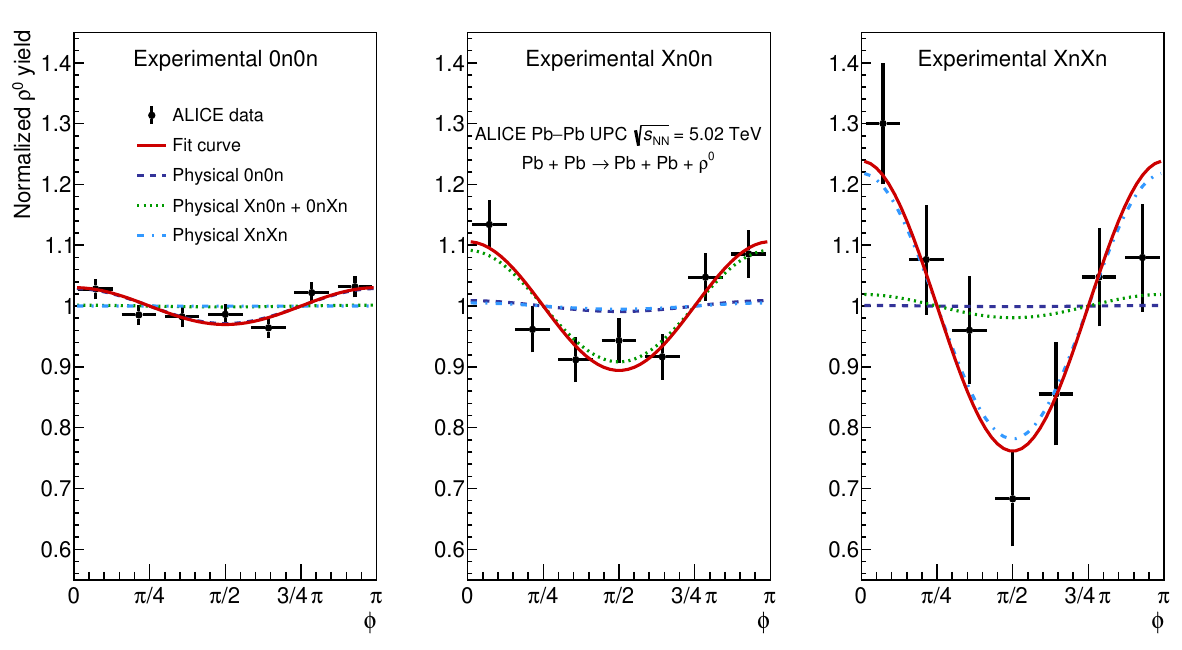}
\caption{Example of a simultaneous fit to the \rhoz yield as a function of \phiz, used to extract the amplitude of the $\cos(2\phi)$ modulation in all neutron classes. The contribution of each physical class to the yield in all experimental classes is shown.
}
\label{Fig:phiFit}  
\end{figure}

The extraction of the amplitude of the modulation is affected by the migration of events between neutron classes, due to ZDC detector efficiency and pile-up effects, as discussed in Ref.~\cite{ALICE:2020ugp}. To take this into account, a simultaneous fit to the measured \rhoz yield as a function of $\phi$ in all three experimental classes (0n0n, Xn0n, XnXn) was performed, using the following expression:
\begin{equation}
    \begin{pmatrix}
         n_{\rho \: \text{0n0n}}(\phi)\\ 
         n_{\rho \: \text{Xn0n}}(\phi)\\
         n_{\rho \: \text{XnXn}}(\phi)
    \end{pmatrix}
    = \begin{pmatrix}
         1 \\ 
         1\\
         1
    \end{pmatrix}
    +
    \begin{pmatrix}
         w_{\text{ 0n0n $\rightarrow$ 0n0n}} & w_{\text{ Xn0n $\rightarrow$ 0n0n}} & w_{\text{ XnXn $\rightarrow$ 0n0n}}\\ 
         w_{\text{ 0n0n $\rightarrow$ Xn0n}} & w_{\text{ Xn0n $\rightarrow$ Xn0n}} & w_{\text{ XnXn $\rightarrow$ Xn0n}}\\ 
         w_{\text{ 0n0n $\rightarrow$ XnXn}} & w_{\text{ Xn0n $\rightarrow$ XnXn}} & w_{\text{ XnXn $\rightarrow$ XnXn}}\\ 
    \end{pmatrix}
    \begin{pmatrix}
         a_{2 \: \text{0n0n}}\\ 
         a_{2 \: \text{Xn0n}}\\ 
         a_{2 \: \text{XnXn}}\\ 
    \end{pmatrix}
     \, \cos(2\phi),
    \label{Eq:phiFitMigrationsMatrix}
\end{equation}
where $n_{\rho \: \text{0n0n}}$ is the normalized \rhoz yield in a given $\phi$ range for the experimental 0n0n class, and similarly for other classes, and the fitting parameters $a_{2 \: \text{0n0n}}$, $a_{2 \: \text{Xn0n}}$, and $a_{2 \: \text{XnXn}}$ are the amplitudes of the $\cos(2\phi)$ modulation in the corresponding three physical classes. The coefficients  $w_{\text{ Y $\rightarrow$ Z}}$ represent the contribution of the physical neutron class~Y to the yield in the experimental neutron class~Z, computed using the measured cross-section ratios and migration probabilities as determined in Ref.~\cite{ALICE:2020ugp}. The constant term is fixed to unity by normalization.  An example of this simultaneous fit is shown in Fig.~\ref{Fig:phiFit}.

The central value and the statistical uncertainty of the cos(2\phiz) modulation amplitude were determined, for each neutron emission class, by averaging the results obtained with the 48 fit configurations described in Sec.~\ref{Sec:signalExtraction}.

\subsection{Systematic uncertainties \label{Sec:syst}}

The systematic uncertainty related to the signal extraction includes the effect on the extracted cos(2$\phi$) amplitudes of variations in the strategy for fitting the invariant mass spectra. These include: binning, range, modeling of the $\rhoz$ signal, and treatment of background. The uncertainty was obtained as the standard deviation of the 
distribution of the fitted amplitudes over the 48 trials mentioned in Sec.~3.4. 
This uncertainty is $12\%$ for 0n0n, $9\%$ for Xn0n, and  $13\%$ for XnXn.

An additional systematic uncertainty, related to the definition of the \phiz angle, was estimated by testing two variations in the analysis strategy. In the first variation, \phiz is computed according to the \textit{charge} definition mentioned in Sec.~\ref{sec:phiDef}. In this case, the yield distribution can have a sizeable $\cos(\phi)$ component~\cite{Hagiwara:2020juc}, which is added to the fit function of Eq.~(\ref{Eq:phiFitMigrationsMatrix}), with its amplitude as an additional free parameter. In the second variation, the \textit{average} definition of $\phiz$ is used, as in the default strategy, but a $\cos(\phi)$ component is nevertheless added to the fit function.  The systematic uncertainty was evaluated in each class as the largest difference between the result obtained with the default setting and that obtained with the two strategies presented in this paragraph. It amounts to $3.6\%$ for 0n0n, $5.6\%$ for Xn0n, and  $3.3\%$ for XnXn.

As a consistency check for the  \axe correction, the analysis was repeated in several rapidity sub-ranges, each containing approximately half the total number of reconstructed \rhoz candidates. In each neutron emission class, the amplitudes extracted in sub-ranges were all found to be compatible, within one standard deviation, with each other and with the amplitude extracted in the full rapidity range. The uncertainty on the \axe correction arises then mainly from the re-weighting procedure described in Sec.~\ref{Sec:corrections}. It was obtained by using the two values of $R_{\text{X}}$ for which the $\chi^2$ increases by one unit with respect to the minimum, instead of the $R_{\text{X}}$ value that minimizes the $\chi^2$. The systematic uncertainty is estimated in each class as the largest difference between the results obtained with the original and with the modified sets of weights. It amounts to $2.9\%$ for 0n0n, $0.8\%$ for Xn0n, and  $0.9\%$ for XnXn.

The systematic uncertainties related to the migration of events across neutron classes are evaluated by propagating the uncertainties of the ZN pile-up probability  
(9\%) and efficiency (1\%), all taken from Ref~\cite{ALICE:2020ugp}, to the extraction of the cos(2$\phi$) amplitude. The resulting uncertainty from pile-up is $0.1\%$, $2.3\%$, and $0.9\%$, respectively, for the 0n0n, Xn0n, and XnXn classes. The uncertainty from the ZN efficiency is  $0.7\%$, $0.1\%$, and $0.1\%$, respectively, for the 0n0n, Xn0n, and XnXn classes.

The contributions to the systematic uncertainty in the amplitude of the $\cos(2\phi$) modulation discussed in the previous paragraphs are listed in Table~\ref{Table:syst}.  The total uncertainty is obtained as the quadratic sum of all the contributions. It amounts to $12.6\%$ for 0n0n, $11\%$ for Xn0n, and  $13.3\%$ for XnXn, and is dominated by the signal extraction for all event classes.

\begin{table}[!ht]
    \centering
    \caption{Summary of the relative systematic uncertainties for the measured amplitude of the $\cos(2\phi)$ modulation.}
    \begin{tabular}{lccc}
    \toprule
    \multirow{2}{*}{Source} & \multicolumn{3}{c}{Uncertainty ($\%$)} \\ 
    & 0n0n & Xn0n~+~0nXn & XnXn \\
    \midrule
    Signal extraction   &   12 & 9.1 & 13 \\
    $\phi$ definition   &   3.6  & 5.7 & 3.3    \\
    \axe                &   2.9  & 0.8 & 0.9  \\
    ZN pile-up          &   0.1  & 2.3 & 0.9  \\
    ZN efficiency       &   0.7  & 0.1 & 0.1  \\
    \midrule
    Total               & 12.6   & 11.0  & 13.3 \\
    \bottomrule
    \label{Table:syst}
    \end{tabular}
\end{table}

\section{Results \label{Sec:results}}
\begin{figure*}[!ht]
\begin{center}
\includegraphics[width=0.96\textwidth]{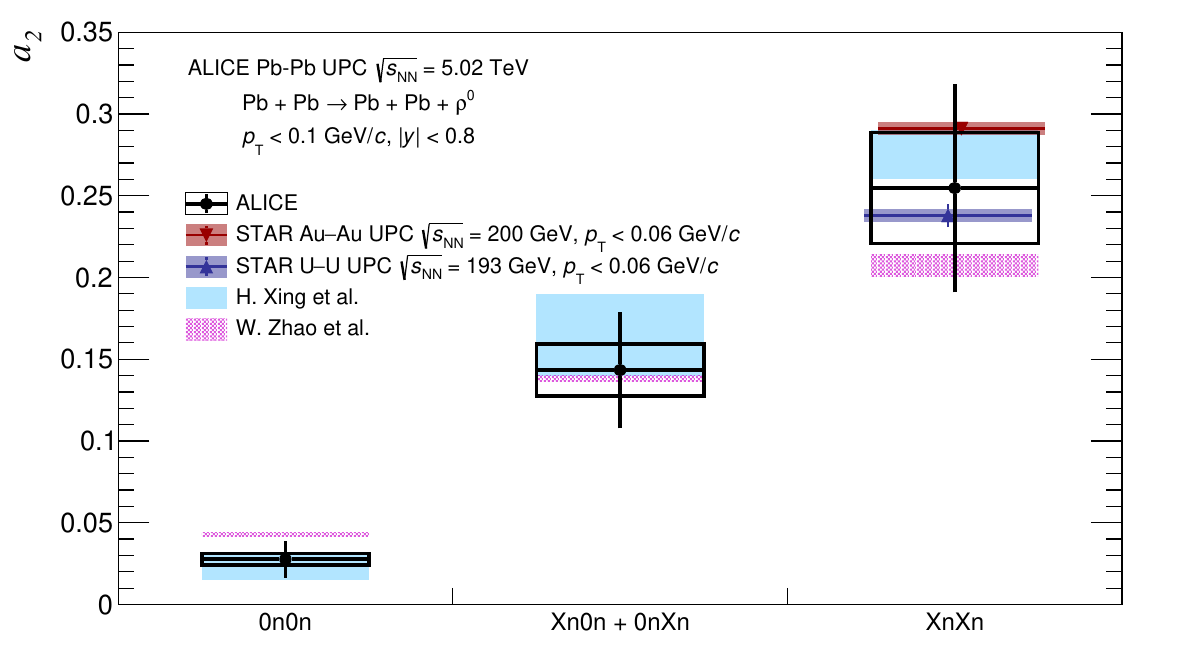}
\caption{Amplitudes of the $\cos(2\phi)$ modulation of the \rhoz yield in Pb--Pb collisions at $\sqrt{s_{\rm{NN}}}$~=~5.02~\TeV in all neutron classes. The results are compared with the Xing \textit{et al.}~\cite{Xing:2020hwh} and W. Zhao \etal~\cite{Mantysaari:2023prg} model predictions and, for the XnXn class, with the STAR results~\cite{STAR:2022wfe} in Au--Au and U--U collisions at RHIC. For all the experimental data points, statistical uncertainties are represented with a bar and systematic uncertainties with a box.}
\label{Fig:modulation}  
\end{center}
\end{figure*}

Figure~\ref{Fig:modulation} shows the extracted amplitude of the $\cos(2\phi)$ modulation as a function of the neutron class; the numerical values are reported in Table~\ref{Table:results}, along with the \Noon~MC estimates of the median impact parameter of the collision for each neutron class. Similar values for the median impact parameters are found using the analytical model presented in Ref.~\cite{Baur:2003ar}; the values for the XnXn case are also similar to those reported in Ref.~\cite{Baltz:2002pp}. The table also reports the amplitudes predicted by two models described below. The measured anisotropy shows a clear trend with the impact parameter, with a significant increase, by one order of magnitude, from 0n0n to XnXn. 

The results are compared with the models by H. Xing \textit{et al.}~\cite{Xing:2020hwh} and by W. Zhao {\em et al.}~\cite{Mantysaari:2023prg}. In both calculations, the \rhoz signal is integrated in the same kinematic region as in this analysis. In the Xing \textit{et al.} model, the quasi-real photon exchanged by the nuclei is treated as a color quark--antiquark dipole, that recombines to produce a \rhoz after scattering off the color-glass-condensate state~\cite{Gelis:2010nm} inside the nuclei. As discussed in Sec.~\ref{sec:intro}, the $\cos(2\phi)$ anisotropy in the model emerges from the presence of two elements: the first is that the photon is linearly polarized along the impact parameter and this polarization is transferred to the produced vector meson; the second is that there is an interference between the two amplitudes that contribute to the cross section of the vector meson photoproduction process. The interference effect increases as the impact parameter decreases, producing a larger anisotropy for small impact parameters.
The uncertainty of the model mostly comes from the probability of emitting a neutron from the scattered nucleus at a given impact parameter, where the latter has been estimated using three different parametrizations from Refs.~\cite{Bertulani:1987tz, Baltz:1996as, Baltz:1997di}. The model prediction is compatible with data for all neutron classes. The model of W. Zhao {\em et al.}~\cite{Mantysaari:2023prg} is based on the same formalism as the model by H. Xing {\em et al.}~\cite{Xing:2020hwh} with two main differences: (\textit{i}) the interaction of the quark--antiquark dipole with the target is implemented by computing the corresponding Wilson lines, and (\textit{ii}) the color-charge density used to obtain the Wilson lines is varied event-by-event to represent the different possible color configurations of the target. The quoted uncertainty in the model originates from the statistical precision related to the finite number of sampled configurations. This model, which predicts a milder variation of the modulation amplitude with the neutron class as compared to Xing {\em et al.}, also gives a reasonable description of the data, with the possible exception of the 0n0n class.

For the XnXn class, the amplitude measured by ALICE is also compared with the ones~\cite{STAR:2022wfe} measured by the STAR Collaboration for \AuAu and \UU collisions at the center-of-mass energies of $\sqrt{s_{\text{NN}}}$~=~200~\GeV and $\sqrt{s_{\text{NN}}}$ = 193~\GeV, respectively. It is found to be compatible with both. This is consistent with the models, which, for the XnXn selection, predict the cos(2$\phi$) modulation amplitude to vary with the colliding nuclei and the center-of-mass energy by less than the current experimental uncertainties.

\begin{table}[!ht]
    \centering
    \caption{Amplitudes of the $\cos(2\phi)$ modulation of the $\rho^0$ yield as a function of $\phi$ in all neutron classes, with statistical and systematic uncertainties. An estimate of the median impact parameter of the collision in each neutron class, obtained with the \Noon~MC, is also reported, as well as the  predictions by the H. Xing~\etal~\cite{Xing:2020hwh} and W. Zhao~\etal~\cite{Mantysaari:2023prg} models.}
    \begin{tabular}{lcccccc}
    \toprule
    Neutron class & median \textit{b} (\Noon) & amplitude & stat. &  syst. &  H. Xing~\etal & W. Zhao~\etal\\
    \midrule
    0n0n & 49.0 fm & 0.028 & 0.011 & 0.003  & 0.015 -- 0.031  & 0.042 -- 0.044 \\ 
    Xn0n & 22.5 fm & 0.14 & 0.04   & 0.016   & 0.14 -- 0.19    & 0.136 -- 0.138 \\
    XnXn & 18.2 fm & 0.25 & 0.06   & 0.03   & 0.26 -- 0.29    & 0.200 -- 0.214 \\
    \bottomrule
    \label{Table:results}
    \end{tabular}
\end{table}

\section{Summary \label{Sec:summary}}
The first measurement of the impact-parameter dependent angular anisotropy in the pion-pair decay of coherently photoproduced $\rho^0$ mesons from Pb–Pb ultraperipheral collisions at a center-of-mass energy of $\sqrt{s_{\text{NN}}}$ = 5.02 TeV, performed with the ALICE detector, has been presented. The anisotropy is quantified via the distribution of the azimuthal angle $\phi$, defined in Sec.~3.2. The impact parameter is estimated considering neutron emission at forward rapidity. 
A significant, impact-parameter dependent $\cos(2\phi)$ modulation is observed, with the amplitude of the modulation increasing by about one order of magnitude from the  0n0n (no neutrons emitted, large impact parameter) to the XnXn (neutrons emitted by both colliding nuclei, relatively small impact parameter) class. This trend is reproduced by
the theoretical models~\cite{Xing:2020hwh,Mantysaari:2023prg}. The result for the XnXn class is compatible with similar measurements by the STAR Collaboration.

The coherent photonuclear production of a vector meson in UPCs can be seen as a double-slit experiment~\cite{Nystrand:2002pd}, where the interference occurs between the amplitudes for quasi-real photon emission by either of the two colliding nuclei. The unambiguous observation of this interference through the measurement of the $\cos(2\phi)$ anisotropy of the \rhoz yield is a proof of the validity of quantum mechanics at  femtometer scales.
This is the first measurement of this effect in terms of the impact-parameter dependence.
The larger data samples expected from the LHC Run 3 and Run 4 will enable a detailed characterization of the quantum interference effects.


\newenvironment{acknowledgement}{\relax}{\relax}
\begin{acknowledgement}
\section*{Acknowledgements}

The ALICE Collaboration would like to thank all its engineers and technicians for their invaluable contributions to the construction of the experiment and the CERN accelerator teams for the outstanding performance of the LHC complex.
The ALICE Collaboration gratefully acknowledges the resources and support provided by all Grid centres and the Worldwide LHC Computing Grid (WLCG) collaboration.
The ALICE Collaboration acknowledges the following funding agencies for their support in building and running the ALICE detector:
A. I. Alikhanyan National Science Laboratory (Yerevan Physics Institute) Foundation (ANSL), State Committee of Science and World Federation of Scientists (WFS), Armenia;
Austrian Academy of Sciences, Austrian Science Fund (FWF): [M 2467-N36] and Nationalstiftung f\"{u}r Forschung, Technologie und Entwicklung, Austria;
Ministry of Communications and High Technologies, National Nuclear Research Center, Azerbaijan;
Conselho Nacional de Desenvolvimento Cient\'{\i}fico e Tecnol\'{o}gico (CNPq), Financiadora de Estudos e Projetos (Finep), Funda\c{c}\~{a}o de Amparo \`{a} Pesquisa do Estado de S\~{a}o Paulo (FAPESP) and Universidade Federal do Rio Grande do Sul (UFRGS), Brazil;
Bulgarian Ministry of Education and Science, within the National Roadmap for Research Infrastructures 2020-2027 (object CERN), Bulgaria;
Ministry of Education of China (MOEC) , Ministry of Science \& Technology of China (MSTC) and National Natural Science Foundation of China (NSFC), China;
Ministry of Science and Education and Croatian Science Foundation, Croatia;
Centro de Aplicaciones Tecnol\'{o}gicas y Desarrollo Nuclear (CEADEN), Cubaenerg\'{\i}a, Cuba;
Ministry of Education, Youth and Sports of the Czech Republic, Czech Republic;
The Danish Council for Independent Research | Natural Sciences, the VILLUM FONDEN and Danish National Research Foundation (DNRF), Denmark;
Helsinki Institute of Physics (HIP), Finland;
Commissariat \`{a} l'Energie Atomique (CEA) and Institut National de Physique Nucl\'{e}aire et de Physique des Particules (IN2P3) and Centre National de la Recherche Scientifique (CNRS), France;
Bundesministerium f\"{u}r Bildung und Forschung (BMBF) and GSI Helmholtzzentrum f\"{u}r Schwerionenforschung GmbH, Germany;
General Secretariat for Research and Technology, Ministry of Education, Research and Religions, Greece;
National Research, Development and Innovation Office, Hungary;
Department of Atomic Energy Government of India (DAE), Department of Science and Technology, Government of India (DST), University Grants Commission, Government of India (UGC) and Council of Scientific and Industrial Research (CSIR), India;
National Research and Innovation Agency - BRIN, Indonesia;
Istituto Nazionale di Fisica Nucleare (INFN), Italy;
Japanese Ministry of Education, Culture, Sports, Science and Technology (MEXT) and Japan Society for the Promotion of Science (JSPS) KAKENHI, Japan;
Consejo Nacional de Ciencia (CONACYT) y Tecnolog\'{i}a, through Fondo de Cooperaci\'{o}n Internacional en Ciencia y Tecnolog\'{i}a (FONCICYT) and Direcci\'{o}n General de Asuntos del Personal Academico (DGAPA), Mexico;
Nederlandse Organisatie voor Wetenschappelijk Onderzoek (NWO), Netherlands;
The Research Council of Norway, Norway;
Pontificia Universidad Cat\'{o}lica del Per\'{u}, Peru;
Ministry of Science and Higher Education, National Science Centre and WUT ID-UB, Poland;
Korea Institute of Science and Technology Information and National Research Foundation of Korea (NRF), Republic of Korea;
Ministry of Education and Scientific Research, Institute of Atomic Physics, Ministry of Research and Innovation and Institute of Atomic Physics and Universitatea Nationala de Stiinta si Tehnologie Politehnica Bucuresti, Romania;
Ministry of Education, Science, Research and Sport of the Slovak Republic, Slovakia;
National Research Foundation of South Africa, South Africa;
Swedish Research Council (VR) and Knut \& Alice Wallenberg Foundation (KAW), Sweden;
European Organization for Nuclear Research, Switzerland;
Suranaree University of Technology (SUT), National Science and Technology Development Agency (NSTDA) and National Science, Research and Innovation Fund (NSRF via PMU-B B05F650021), Thailand;
Turkish Energy, Nuclear and Mineral Research Agency (TENMAK), Turkey;
National Academy of  Sciences of Ukraine, Ukraine;
Science and Technology Facilities Council (STFC), United Kingdom;
National Science Foundation of the United States of America (NSF) and United States Department of Energy, Office of Nuclear Physics (DOE NP), United States of America.
In addition, individual groups or members have received support from:
Czech Science Foundation (grant no. 23-07499S), Czech Republic;
European Research Council (grant no. 950692), European Union;
ICSC - Centro Nazionale di Ricerca in High Performance Computing, Big Data and Quantum Computing, European Union - NextGenerationEU;
Academy of Finland (Center of Excellence in Quark Matter) (grant nos. 346327, 346328), Finland.

\end{acknowledgement}

\bibliographystyle{utphys}   
\bibliography{bibliography}

\newpage
\appendix

%
%

\section{The ALICE Collaboration}
\label{app:collab}
\begin{flushleft} 
\small

S.~Acharya\,\orcidlink{0000-0002-9213-5329}\,$^{\rm 127}$, 
D.~Adamov\'{a}\,\orcidlink{0000-0002-0504-7428}\,$^{\rm 86}$, 
A.~Agarwal$^{\rm 135}$, 
G.~Aglieri Rinella\,\orcidlink{0000-0002-9611-3696}\,$^{\rm 32}$, 
L.~Aglietta\,\orcidlink{0009-0003-0763-6802}\,$^{\rm 24}$, 
M.~Agnello\,\orcidlink{0000-0002-0760-5075}\,$^{\rm 29}$, 
N.~Agrawal\,\orcidlink{0000-0003-0348-9836}\,$^{\rm 25}$, 
Z.~Ahammed\,\orcidlink{0000-0001-5241-7412}\,$^{\rm 135}$, 
S.~Ahmad\,\orcidlink{0000-0003-0497-5705}\,$^{\rm 15}$, 
S.U.~Ahn\,\orcidlink{0000-0001-8847-489X}\,$^{\rm 71}$
% ...

\section*{Affiliation Notes}

$^{\rm I}$ Deceased\\
$^{\rm II}$ Also at: Max-Planck-Institut fur Physik, Munich, Germany\\
$^{\rm III}$ Also at: Italian National Agency for New Technologies, Energy and Sustainable Economic Development (ENEA), Bologna, Italy\\
$^{\rm IV}$ Also at: Dipartimento DET del Politecnico di Torino, Turin, Italy\\
$^{\rm V}$ Also at: Yildiz Technical University, Istanbul, T\"{u}rkiye\\
$^{\rm VI}$ Also at: Department of Applied Physics, Aligarh Muslim University, Aligarh, India\\
$^{\rm VII}$ Also at: Institute of Theoretical Physics, University of Wroclaw, Poland\\
$^{\rm VIII}$ Also at: An institution covered by a cooperation agreement with CERN\\

\section*{Collaboration Institutes}

$^{1}$ A.I. Alikhanyan National Science Laboratory (Yerevan Physics Institute) Foundation, Yerevan, Armenia\\
$^{2}$ AGH University of Krakow, Cracow, Poland\\
$^{3}$ Bogolyubov Institute for Theoretical Physics, National Academy of Sciences of Ukraine, Kiev, Ukraine\\
$^{4}$ Bose Institute, Department of Physics  and Centre for Astroparticle Physics and Space Science (CAPSS), Kolkata, India\\
$^{5}$ California Polytechnic State University, San Luis Obispo, California, United States\\
$^{6}$ Central China Normal University, Wuhan, China\\
$^{7}$ Centro de Aplicaciones Tecnol\'{o}gicas y Desarrollo Nuclear (CEADEN), Havana, Cuba\\
$^{8}$ Centro de Investigaci\'{o}n y de Estudios Avanzados (CINVESTAV), Mexico City and M\'{e}rida, Mexico\\
$^{9}$ Chicago State University, Chicago, Illinois, United States\\
$^{10}$ China Institute of Atomic Energy, Beijing, China\\
$^{11}$ China University of Geosciences, Wuhan, China\\
$^{12}$ Chungbuk National University, Cheongju, Republic of Korea\\
$^{13}$ Comenius University Bratislava, Faculty of Mathematics, Physics and Informatics, Bratislava, Slovak Republic\\
$^{14}$ Creighton University, Omaha, Nebraska, United States\\
$^{15}$ Department of Physics, Aligarh Muslim University, Aligarh, India\\
$^{16}$ Department of Physics, Pusan National University, Pusan, Republic of Korea\\
$^{17}$ Department of Physics, Sejong University, Seoul, Republic of Korea\\
$^{18}$ Department of Physics, University of California, Berkeley, California, United States\\
$^{19}$ Department of Physics, University of Oslo, Oslo, Norway\\
$^{20}$ Department of Physics and Technology, University of Bergen, Bergen, Norway\\
$^{21}$ Dipartimento di Fisica, Universit\`{a} di Pavia, Pavia, Italy\\
$^{22}$ Dipartimento di Fisica dell'Universit\`{a} and Sezione INFN, Cagliari, Italy\\
$^{23}$ Dipartimento di Fisica dell'Universit\`{a} and Sezione INFN, Trieste, Italy\\
$^{24}$ Dipartimento di Fisica dell'Universit\`{a} and Sezione INFN, Turin, Italy\\
$^{25}$ Dipartimento di Fisica e Astronomia dell'Universit\`{a} and Sezione INFN, Bologna, Italy\\
$^{26}$ Dipartimento di Fisica e Astronomia dell'Universit\`{a} and Sezione INFN, Catania, Italy\\
$^{27}$ Dipartimento di Fisica e Astronomia dell'Universit\`{a} and Sezione INFN, Padova, Italy\\
$^{28}$ Dipartimento di Fisica `E.R.~Caianiello' dell'Universit\`{a} and Gruppo Collegato INFN, Salerno, Italy\\
$^{29}$ Dipartimento DISAT del Politecnico and Sezione INFN, Turin, Italy\\
$^{30}$ Dipartimento di Scienze MIFT, Universit\`{a} di Messina, Messina, Italy\\
$^{31}$ Dipartimento Interateneo di Fisica `M.~Merlin' and Sezione INFN, Bari, Italy\\
$^{32}$ European Organization for Nuclear Research (CERN), Geneva, Switzerland\\
$^{33}$ Faculty of Electrical Engineering, Mechanical Engineering and Naval Architecture, University of Split, Split, Croatia\\
$^{34}$ Faculty of Engineering and Science, Western Norway University of Applied Sciences, Bergen, Norway\\
$^{35}$ Faculty of Nuclear Sciences and Physical Engineering, Czech Technical University in Prague, Prague, Czech Republic\\
$^{36}$ Faculty of Physics, Sofia University, Sofia, Bulgaria\\
$^{37}$ Faculty of Science, P.J.~\v{S}af\'{a}rik University, Ko\v{s}ice, Slovak Republic\\
$^{38}$ Frankfurt Institute for Advanced Studies, Johann Wolfgang Goethe-Universit\"{a}t Frankfurt, Frankfurt, Germany\\
$^{39}$ Fudan University, Shanghai, China\\
$^{40}$ Gangneung-Wonju National University, Gangneung, Republic of Korea\\
$^{41}$ Gauhati University, Department of Physics, Guwahati, India\\
$^{42}$ Helmholtz-Institut f\"{u}r Strahlen- und Kernphysik, Rheinische Friedrich-Wilhelms-Universit\"{a}t Bonn, Bonn, Germany\\
$^{43}$ Helsinki Institute of Physics (HIP), Helsinki, Finland\\
$^{44}$ High Energy Physics Group,  Universidad Aut\'{o}noma de Puebla, Puebla, Mexico\\
$^{45}$ Horia Hulubei National Institute of Physics and Nuclear Engineering, Bucharest, Romania\\
$^{46}$ HUN-REN Wigner Research Centre for Physics, Budapest, Hungary\\
$^{47}$ Indian Institute of Technology Bombay (IIT), Mumbai, India\\
$^{48}$ Indian Institute of Technology Indore, Indore, India\\
$^{49}$ INFN, Laboratori Nazionali di Frascati, Frascati, Italy\\
$^{50}$ INFN, Sezione di Bari, Bari, Italy\\
$^{51}$ INFN, Sezione di Bologna, Bologna, Italy\\
$^{52}$ INFN, Sezione di Cagliari, Cagliari, Italy\\
$^{53}$ INFN, Sezione di Catania, Catania, Italy\\
$^{54}$ INFN, Sezione di Padova, Padova, Italy\\
$^{55}$ INFN, Sezione di Pavia, Pavia, Italy\\
$^{56}$ INFN, Sezione di Torino, Turin, Italy\\
$^{57}$ INFN, Sezione di Trieste, Trieste, Italy\\
$^{58}$ Inha University, Incheon, Republic of Korea\\
$^{59}$ Institute for Gravitational and Subatomic Physics (GRASP), Utrecht University/Nikhef, Utrecht, Netherlands\\
$^{60}$ Institute of Experimental Physics, Slovak Academy of Sciences, Ko\v{s}ice, Slovak Republic\\
$^{61}$ Institute of Physics, Homi Bhabha National Institute, Bhubaneswar, India\\
$^{62}$ Institute of Physics of the Czech Academy of Sciences, Prague, Czech Republic\\
$^{63}$ Institute of Space Science (ISS), Bucharest, Romania\\
$^{64}$ Institut f\"{u}r Kernphysik, Johann Wolfgang Goethe-Universit\"{a}t Frankfurt, Frankfurt, Germany\\
$^{65}$ Instituto de Ciencias Nucleares, Universidad Nacional Aut\'{o}noma de M\'{e}xico, Mexico City, Mexico\\
$^{66}$ Instituto de F\'{i}sica, Universidade Federal do Rio Grande do Sul (UFRGS), Porto Alegre, Brazil\\
$^{67}$ Instituto de F\'{\i}sica, Universidad Nacional Aut\'{o}noma de M\'{e}xico, Mexico City, Mexico\\
$^{68}$ iThemba LABS, National Research Foundation, Somerset West, South Africa\\
$^{69}$ Jeonbuk National University, Jeonju, Republic of Korea\\
$^{70}$ Johann-Wolfgang-Goethe Universit\"{a}t Frankfurt Institut f\"{u}r Informatik, Fachbereich Informatik und Mathematik, Frankfurt, Germany\\
$^{71}$ Korea Institute of Science and Technology Information, Daejeon, Republic of Korea\\
$^{72}$ KTO Karatay University, Konya, Turkey\\
$^{73}$ Laboratoire de Physique Subatomique et de Cosmologie, Universit\'{e} Grenoble-Alpes, CNRS-IN2P3, Grenoble, France\\
$^{74}$ Lawrence Berkeley National Laboratory, Berkeley, California, United States\\
$^{75}$ Lund University Department of Physics, Division of Particle Physics, Lund, Sweden\\
$^{76}$ Nagasaki Institute of Applied Science, Nagasaki, Japan\\
$^{77}$ Nara Women{'}s University (NWU), Nara, Japan\\
$^{78}$ National and Kapodistrian University of Athens, School of Science, Department of Physics , Athens, Greece\\
$^{79}$ National Centre for Nuclear Research, Warsaw, Poland\\
$^{80}$ National Institute of Science Education and Research, Homi Bhabha National Institute, Jatni, India\\
$^{81}$ National Nuclear Research Center, Baku, Azerbaijan\\
$^{82}$ National Research and Innovation Agency - BRIN, Jakarta, Indonesia\\
$^{83}$ Niels Bohr Institute, University of Copenhagen, Copenhagen, Denmark\\
$^{84}$ Nikhef, National institute for subatomic physics, Amsterdam, Netherlands\\
$^{85}$ Nuclear Physics Group, STFC Daresbury Laboratory, Daresbury, United Kingdom\\
$^{86}$ Nuclear Physics Institute of the Czech Academy of Sciences, Husinec-\v{R}e\v{z}, Czech Republic\\
$^{87}$ Oak Ridge National Laboratory, Oak Ridge, Tennessee, United States\\
$^{88}$ Ohio State University, Columbus, Ohio, United States\\
$^{89}$ Physics department, Faculty of science, University of Zagreb, Zagreb, Croatia\\
$^{90}$ Physics Department, Panjab University, Chandigarh, India\\
$^{91}$ Physics Department, University of Jammu, Jammu, India\\
$^{92}$ Physics Program and International Institute for Sustainability with Knotted Chiral Meta Matter (SKCM2), Hiroshima University, Hiroshima, Japan\\
$^{93}$ Physikalisches Institut, Eberhard-Karls-Universit\"{a}t T\"{u}bingen, T\"{u}bingen, Germany\\
$^{94}$ Physikalisches Institut, Ruprecht-Karls-Universit\"{a}t Heidelberg, Heidelberg, Germany\\
$^{95}$ Physik Department, Technische Universit\"{a}t M\"{u}nchen, Munich, Germany\\
$^{96}$ Politecnico di Bari and Sezione INFN, Bari, Italy\\
$^{97}$ Research Division and ExtreMe Matter Institute EMMI, GSI Helmholtzzentrum f\"ur Schwerionenforschung GmbH, Darmstadt, Germany\\
$^{98}$ Saga University, Saga, Japan\\
$^{99}$ Saha Institute of Nuclear Physics, Homi Bhabha National Institute, Kolkata, India\\
$^{100}$ School of Physics and Astronomy, University of Birmingham, Birmingham, United Kingdom\\
$^{101}$ Secci\'{o}n F\'{\i}sica, Departamento de Ciencias, Pontificia Universidad Cat\'{o}lica del Per\'{u}, Lima, Peru\\
$^{102}$ Stefan Meyer Institut f\"{u}r Subatomare Physik (SMI), Vienna, Austria\\
$^{103}$ SUBATECH, IMT Atlantique, Nantes Universit\'{e}, CNRS-IN2P3, Nantes, France\\
$^{104}$ Sungkyunkwan University, Suwon City, Republic of Korea\\
$^{105}$ Suranaree University of Technology, Nakhon Ratchasima, Thailand\\
$^{106}$ Technical University of Ko\v{s}ice, Ko\v{s}ice, Slovak Republic\\
$^{107}$ The Henryk Niewodniczanski Institute of Nuclear Physics, Polish Academy of Sciences, Cracow, Poland\\
$^{108}$ The University of Texas at Austin, Austin, Texas, United States\\
$^{109}$ Universidad Aut\'{o}noma de Sinaloa, Culiac\'{a}n, Mexico\\
$^{110}$ Universidade de S\~{a}o Paulo (USP), S\~{a}o Paulo, Brazil\\
$^{111}$ Universidade Estadual de Campinas (UNICAMP), Campinas, Brazil\\
$^{112}$ Universidade Federal do ABC, Santo Andre, Brazil\\
$^{113}$ Universitatea Nationala de Stiinta si Tehnologie Politehnica Bucuresti, Bucharest, Romania\\
$^{114}$ University of Cape Town, Cape Town, South Africa\\
$^{115}$ University of Derby, Derby, United Kingdom\\
$^{116}$ University of Houston, Houston, Texas, United States\\
$^{117}$ University of Jyv\"{a}skyl\"{a}, Jyv\"{a}skyl\"{a}, Finland\\
$^{118}$ University of Kansas, Lawrence, Kansas, United States\\
$^{119}$ University of Liverpool, Liverpool, United Kingdom\\
$^{120}$ University of Science and Technology of China, Hefei, China\\
$^{121}$ University of South-Eastern Norway, Kongsberg, Norway\\
$^{122}$ University of Tennessee, Knoxville, Tennessee, United States\\
$^{123}$ University of the Witwatersrand, Johannesburg, South Africa\\
$^{124}$ University of Tokyo, Tokyo, Japan\\
$^{125}$ University of Tsukuba, Tsukuba, Japan\\
$^{126}$ Universit\"{a}t M\"{u}nster, Institut f\"{u}r Kernphysik, M\"{u}nster, Germany\\
$^{127}$ Universit\'{e} Clermont Auvergne, CNRS/IN2P3, LPC, Clermont-Ferrand, France\\
$^{128}$ Universit\'{e} de Lyon, CNRS/IN2P3, Institut de Physique des 2 Infinis de Lyon, Lyon, France\\
$^{129}$ Universit\'{e} de Strasbourg, CNRS, IPHC UMR 7178, F-67000 Strasbourg, France, Strasbourg, France\\
$^{130}$ Universit\'{e} Paris-Saclay, Centre d'Etudes de Saclay (CEA), IRFU, D\'{e}partment de Physique Nucl\'{e}aire (DPhN), Saclay, France\\
$^{131}$ Universit\'{e}  Paris-Saclay, CNRS/IN2P3, IJCLab, Orsay, France\\
$^{132}$ Universit\`{a} degli Studi di Foggia, Foggia, Italy\\
$^{133}$ Universit\`{a} del Piemonte Orientale, Vercelli, Italy\\
$^{134}$ Universit\`{a} di Brescia, Brescia, Italy\\
$^{135}$ Variable Energy Cyclotron Centre, Homi Bhabha National Institute, Kolkata, India\\
$^{136}$ Warsaw University of Technology, Warsaw, Poland\\
$^{137}$ Wayne State University, Detroit, Michigan, United States\\
$^{138}$ Yale University, New Haven, Connecticut, United States\\
$^{139}$ Yonsei University, Seoul, Republic of Korea\\
$^{140}$  Zentrum  f\"{u}r Technologie und Transfer (ZTT), Worms, Germany\\
$^{141}$ Affiliated with an institute covered by a cooperation agreement with CERN\\
$^{142}$ Affiliated with an international laboratory covered by a cooperation agreement with CERN.\\

\end{flushleft} 

\end{document}